\documentstyle[epsf]{article}
\begin{document}

\parskip 2mm plus 1mm \parindent=0pt
\def\cl{\centerline}
 \def\hs1{\hskip1mm}\def\hr{\hskip.2mm} 
\def\h10{\hskip10mm}\def\hsp{\hskip.5mm} \def\page{\vfill\eject}
\def\hx{\hskip10mm\hbox} \def\vs{\vskip4mm}
\def\<{\langle} \def\>{\rangle} \def\br{\bf\rm}  
 \def\de{\partial} \def\Tr{{\rm Tr}}\def\dag{^\dagger} 
\def\sadj{\hskip1mm{\rm sadj}}

\def\ne{=\hskip-3.3mm /\hskip3.3mm} 
\def\half{{\scriptstyle{1\over 2}}} \def\quter{{\scriptstyle{1\over 4}}} 
\def\Im{{\rm Im}} \def\Re{{\rm Re}} \def\I{{\rm I}} 
 \def\R{{\rm R}}  \def\M{{\br M}\hskip0.7mm} 
\def\G{{\bf G}}\def\F{{\bf F}}\def\H{{\bf H}}
 \def\P{{\rm P}} \def\Pr{{\br Pr}} \def\cc{{\rm cc}} \def\HC{{\rm
HC}} \def\D{{\br D}} \def\N{{\br N}} \def\S{{\rm S}}
\def\x{{\rm x}} \def\d{{\rm d}}\def\i{{\rm i}}\def\f{{\rm f}} 
\def\La{L_{\rm a}}\def\Lb{L_{\rm b}}\def\Lc{L_{\rm c}} 
 \def\Sl{S_{\rm loc}}
\def\Sc{S_{\rm c}} \def\ab{\overline a} 
\def\xv{\vec x}

\def\CL{\longrightarrow^{\hskip-5mm \rm CL}\hs1}
\def\NI{\longrightarrow^{\hskip-5mm \rm NI}\hs1}

\vs\bf\cl{Cosmic quantum measurement} 

\vs

\centerline {by}

\vs

\centerline {I. C. Percival}\vs 

\centerline {Department of Physics} 
\centerline {Queen Mary and Westfield College, University of London} 
\centerline {Mile End Road, London E1 4NS, England}

\vskip15mm \centerline{\bf Abstract} \rm 

Hardy's theorem states that the hidden variables of any realistic
theory of quantum measurement, whose predictions agree with ordinary
quantum theory, must have a preferred Lorentz frame.  This presents
the conflict between special relativity and any realistic dynamics of
quantum measurement in a severe form.  The conflict is resolved using
a `measurement field', which provides a timelike function of spacetime
points and a definition of simultaneity in the context of a curved
spacetime.  Locally this theory is consistent with special relativity,
but globally, special relativity is not enough; the time dilation of
general relativity and the standard cosmic time of the
Robertson-Walker cosmologies are both essential.  A simple but crude
example is a relativistic quantum measurement dynamics based on the
nonrelativistic measurement dynamics of L\"uders.

\vskip55mm
\vfill
 98 Nov 30, QMW-PH-98-??\hfill Submitted to ??
\page

\vskip10mm

{\bf 1 Introduction}

{\bf 2 Hardy's theorem} 

{\bf 3 Simultaneity}

{\bf 4 Not proper time}

{\bf 5 Cosmology}

{\bf 6 Measurement field}

{\bf 7 A simple relativistic measurement dynamics}

{\bf 8 Discussion and conclusions}


\vs
\vs{\bf 1 Introduction} 

The theoretical ideas of Hardy \cite{Hardy1992a}, Bell
\cite{Bell1987a}, Wheeler \cite{Wheeler1983b} and Hawking and Seifert
\cite{Hawking1968,Tipler1980}, combined with the long-range
entanglement experiments of Tittel, Gisin and collaborators
\cite{Tittel1998a,Tittel1998b}, lead to an elementary cosmological
dynamics of quantum measurement that satisfies the principles of
special and general relativity.

According to Bohr \cite{Bohr1935,Wheeler1983a}, the result of a
quantum measurement is influenced by the condition of the measuring
apparatus.  Dynamical theories of quantum measurement ascribe this
influence to a dynamical process.  Here the influence is due to
interaction with a cosmic background field, the {\it measurement
field} which plays the role of hidden variables in some other
dynamical theories.  By an extension of the Einstein, Podolsky and
Rosen thought experiment \cite{Einstein1935}, John Bell showed
\cite{Bell1990a}, \cite{Bell1990b} that quantum theories which
represent quantum measurement as a dynamical process are nonlocal.
Hardy \cite{Hardy1992a} showed the need for a preferred Lorentz frame.
The measurement field provides the frame.

{\it Generalized} quantum measurement is a physical process by
which the state of a quantum system influences the value of a
classical variable.  It includes any such process, for example
laboratory measurements, but also other, very different, processes
\cite{Percival1998c}.  These include the cosmic rays
that produced small but detectable dislocations in mineral crystals
during the Jurassic era, and the quantum fluctuations in the early
universe that may have caused today's anisotropies in the universe.
It includes those quantum fluctuations that are amplified by chaotic
dynamics to produce significant changes in classical dynamical
variables.  For such measurements, these are the dynamical variables
of the {\it measurer}, although there is no measuring apparatus in the
usual sense.

Assume, therefore, that quantum measurement is universal, that the
measurement process is taking place throughout spacetime, with the
possible exception of the neighbourhood of some spacetime
singularities.  Here I also assume that quantum measurement dynamics
includes a specific representation for the evolution of an individual
quantum system: the evolution of an ensemble is not enough.  And
assume that there are no causal loops in spacetime.

This picture and the results that follow are based on the following
five principles PR1-5 and two theorems PR6,7:

\vs 
\hspace{1cm}\parbox{11cm}{
PR1. {\it Kepler-Galileo} - Humans are not at the centre of the
	universe, in any sense.

PR2. {\it Newton-Laplace} - Every physical process has a dynamical explanation.

PR3. {\it Einstein} - Special relativity.

PR4. {\it Einstein} - General relativity.

PR5. {\it Cosmological} - On sufficiently large scales the universe is
spatially isotropic and therefore uniform.

PR6. {\it Hardy \cite{Hardy1992a}} - Measurement dynamics in flat
spacetime and consistent with ordinary quantum theory needs a special
Lorentz frame.

PR7. {\it Hardy simultaneity} - Measurement dynamics in curved
spacetime needs a definition of simultaneity for events with spacelike
separation.}

\vs 

I also assume that
The principles PR1 and PR2 together are incompatible with ordinary
quantum mechanics, in which quantum measurement is the preserve of
humans, and does not require a dynamical explanation.  Despite this,
it is our most successful theory.  A good reason for the recent
revival of alternative realistic theories which are compatible with
these principles is the greater control we have over quantum systems,
which raises the possibility of distinguishing different quantum
theories of measurement experimentally.  Notice that the (weak)
version of the Newton-Laplace principle PR2 does not require the
dynamics to be deterministic.

Aharanov and Albert \cite{Aharanov1981,Aharanov1984a,Aharanov1984b}
pointed to the particular difficulties of reconciling special
relativity and realistic quantum theories.  According to Shimony
\cite{Shimony1986}, special relativity and quantum mechanics might
live in `peaceful coexistence', but Hardy's theorem suggests a
fundamental conflict between the expected results of quantum
measurement and invariance under Lorentz transformations.  The
principal purpose of this paper is to resolve this apparent conflict
between special relativity and the dynamics of quantum measurement.

There are now many alternative quantum theories that provide a
nonrelativistic dynamics of quantum measurement.  To my knowledge,
there has been no alternative theory that resolves the major problem
of reconciling special relativity and the dynamics of quantum
measurement in general, or Hardy's theorem PR6 in particular.  Nor
have I been able to formulate a consistent relativistic dynamical
theory of quantum measurement that applies to our universe without
including both general relativity and cosmology.

In order to make the paper more accessible, sections 2 to 4 include
short reviews of the relevant quantum theory for general relativists
and cosmologists, and some relevant general relativity and cosmology
for quantum theorists.

Section 2 describes Bell's theorem and Hardy's theorem.  Section 3
presents an alternative proof of Hardy's theorem based on a
combination of two classically connected Bell experiments.  Their
spacetime configuration is similar to that of the no-simultaneity
thought experiment of Einstein's original work on special relativity.
It goes on to sketch a proof the Hardy simultaneity theorem, which is
an extension of his original theorem, but for curved spacetime, and to
state the Hawking-Seifert theorem on timelike functions in some
general curved spacetimes.

Section 4 uses general relativistic time dilation to demonstrate that
for two clocks at different heights on the Earth or at different
locations in the universe, local proper times do not provide a global
definition of simultaneity.  Two reasons for a cosmological theory of
quantum measurement are given in section 5, and section 6
introduces the measurement field, which provides the simultaneity
needed for such a theory, and for the resolution of the conflict
between quantum measurement and relativity.

Section 7 sketches a simple example of relativistic measurement
dynamics based on the measurement field, which leaves much room for
improvement, and the final section 8 includes a brief discussion
of the relation between this dynamics and some other alternative
quantum theories.

\vs{\bf 2 Hardy's theorem} 

Hardy's theorem PR6 \cite{Hardy1992a} goes further than Bell's theorem
on nonlocality of quantum measurement.  From this theorem it follows
that any dynamical theory of measurement, in which the results of the
measurements agree with those of ordinary quantum theory, must have a
preferred Lorentz frame.  The theorem does not determine this frame.

Bell's theorem shows that measurement dynamics is nonlocal if the
results of measurements follow the rules of ordinary quantum theory
\cite{Bell1964,Bell1987a}.  Bell demonstrated his theorem by a thought
experiment illustrated in figure 1,
\begin{figure}[htb]
\begin{center}
\epsfxsize14.0cm
\centerline{\epsfbox{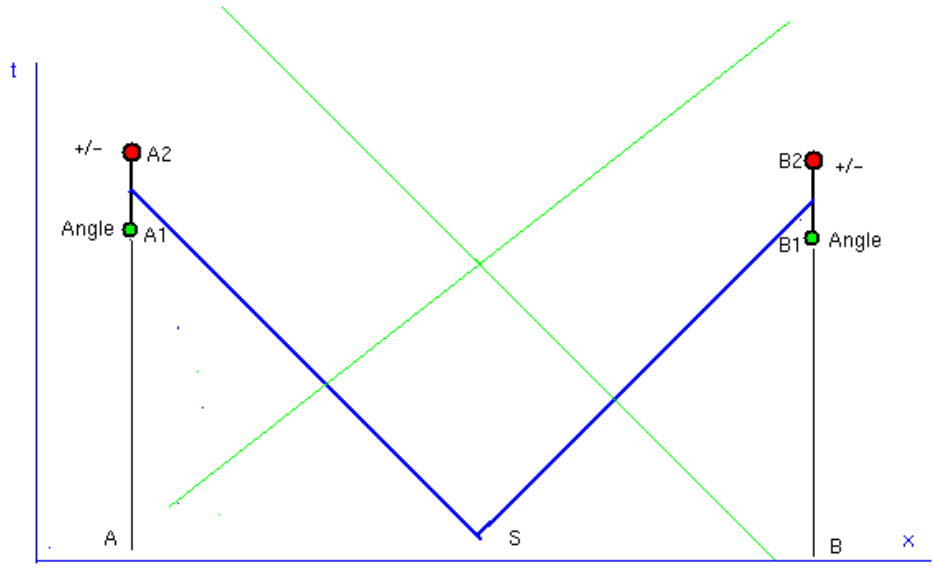} }
\label{fig1}
\end{center}
\caption{Spacetime diagram of Bell's experiment. Thin diagonal lines
at $45^o$ represent the velocity of light.  At both A1 and B1, the
angle is the setting of the angle of spin or polarization measurement,
which is an input event, and at A2 and B2, $+/-$ represents the recording of
the spin or polarization, an output event.}
\end{figure}
in which two entangled particles, each with spin nonzero, and with
total spin zero, are produced from a source S.  A component of spin
perpendicular to the direction of motion of each particle is measured,
one at A, and the other at B, where typically ASB is a straight line,
with S at its centre.  The alignment of the spin measuring apparatus
at A or B is the preparation event, or input, A1 or B1, and the
measurement and recording of the spin component is the measurement
event, or output, A2 or B2.  Both the events A1,A2 at A have spacelike
separation from both the events B1,B2 at B.  In the illustrated
example the particles are photons, and a line at $45\deg$ represents a
photon at the velocity of light.  Bell's theorem then states that for
any realistic dynamics of quantum measurement, if the results agree
with ordinary quantum theory, either the input at A1 affects the
output at B2 or the input at B1 affects the output at A2, or both.
There must be nonlocal space-like causality.

Such an experiment was carried out by Aspect and his collaborators
\cite{Aspect1982a,Aspect1982b,Peres1995a}, although there remains at
least one loophole to be closed because, despite the considerable care
that was taken, it is not clear that the events fully satisfied the
spacelike separation condition.  \cite{Fry1997,Walther1997}.

Hardy demonstrates his theorem through a thought experiment involving
two matter interferometers, one for electrons and one for positrons,
with an intersection between them that allows annihilation of the
particles to produce gamma rays.  In its original form, this
experiment is likely to remain a thought experiment.  Improved
versions which depend on similar principles and are experimentally
feasible are given in \cite{Clifton1992,Hardy1992c}.  A different
derivation based on classical links between two Bell experiments is
given in \cite{Percival1998b} and section 3.

\vs{\bf 3 Simultaneity}

In classical special relativity with flat spacetimes there is no
unique simultaneity for events with spacelike separation.  This was
demonstrated by Einstein in the famous classical thought experiment,
which we describe for later convenience.  In Einstein's experiment,
illustrated in figure 2, 
\begin{figure}[htb]
\begin{center}
\epsfxsize14.0cm
\centerline{\epsfbox{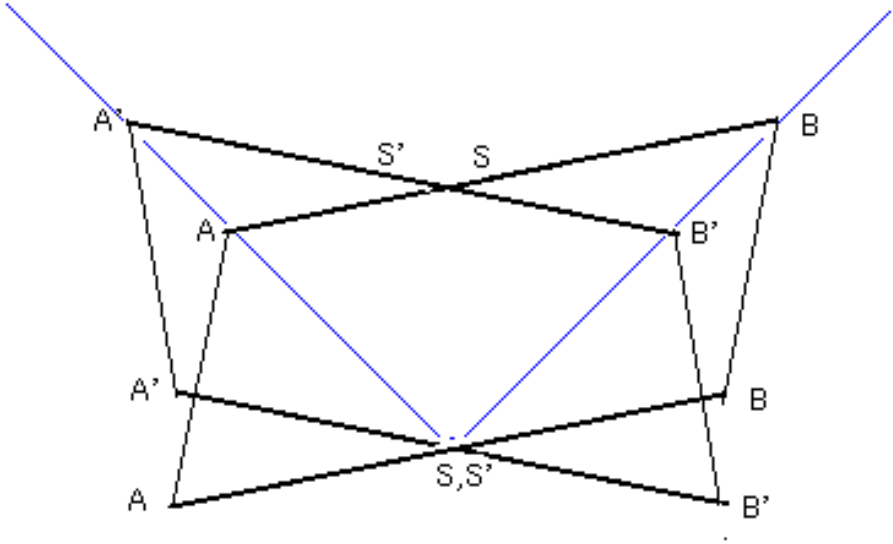} }
\label{fig2}
\end{center}
\caption{Spacetime diagram of Einstein's simultaneity experiment.  The
straight lines ASB and A$'$S$'$B$'$ represent a source and two receivers
at rest in their respective frames.}
\end{figure}

one part consists of a source S which emits a flash of light, and two
receivers A and B, equidistant from S in the same straight line, where
A,S and B are at rest in a frame L.  It is received simultaneously at
A and B with respect to this frame.  The other part of the experiment
consists of an identical trio A$'$S$'$B$'$, in the same straight line
as ASB, which are at rest in a different frame L$'$, moving with
respect to L in the direction AOB, where S and S$'$ are nearly
coincident at the time $t_0$ when both of them flash.  The light from
S$'$ is received simultaneously at A$'$ and B$'$ with respect to L$'$,
and this is clearly not simultaneous with respect to L.  Hence
relativity.

There is an alternative proof \cite{Percival1998b} of Hardy's theorem,
which depends on two Bell experiments in a similar spacetime
configuration to Einstein's simultaneity experiment, and labelled
similarly in figure 3.  This is the double Bell experiment.  
\begin{figure}[htb]
\begin{center}
\epsfxsize14.0cm
\centerline{\epsfbox{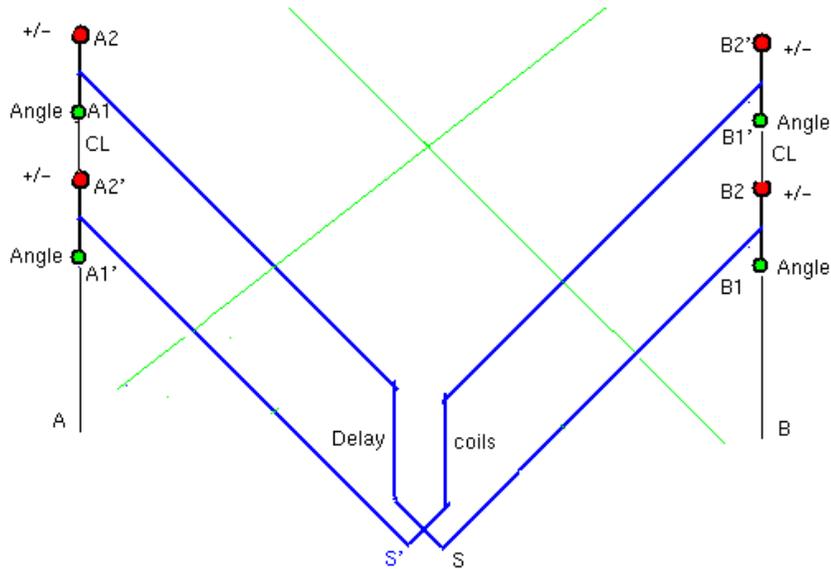} }
\label{fig3}
\end{center}
\caption{Spacetime diagram of the double Bell experiment, with photons
in optical fibres.  The meaning of the symbols corresponds to their
meaning in the first two figures. On the scale of this figure, the
photons remain at rest while in the delay coils.  CL are classical
links, by which an output from each experiment determines an input to
the other.}
\end{figure}
The two Bell experiments are independent at the quantum level, but
they are linked classically so that the output A2 controls the
input A1$'$, which is in its future lightcone, and the output B2$'$
controls the input B1.  According to Bell's theorem, there are
nonlocal interactions NI, such as the setting of the angle at A1 affecting
the measurement at B2, or the setting of the angle at B1' affecting
the measurement at A2'.  If {\it both} of these are present, then there is
a causal loop
\begin{equation}
{\rm A2}'\CL{\rm A1}\NI{\rm B2}\CL{\rm B1}'\NI{\rm A2}'
\end{equation}

The assumption that there are no causal loops in spacetime or the
equivalent assumption that there is no backward causality, then leads
to the exclusion of one of the nonlocal interactions, which
makes the dynamics dependent on the Lorentz frame, and so leads to
Hardy's theorem.  Details are in the letter [28]. 

As it stands, Hardy's theorem makes no statement about simultaneity,
but this also can be derived using the double Bell experiment.
Thus a configuration similary to that used by Einstein to show that
classical special relativity has no universal simultaneity, can be
used to show that quantum measurement {\it requires} universal
simultaneity, which is the Hardy simultaneity theorem PR7 of the
introduction.

In flat spacetimes, a preferred Lorentz frame or a standard of rest
defines a universal time and simultaneity between distant events.  For
curved spacetimes, only a local Lorentz frame or standard of rest has
meaning, and even if there is a standard of rest for every point, this
does not necessarily provide a definition of simultaneity.  The
examples of the next section show that in the presence of
gravitational fields, the local times defined by local preferred
Lorentz frames are not necessarily consistent with a universal
simultaneity defined throughout a region.  For curved spacetimes,
simultaneity is a stronger condition.  For flat spacetimes they are
equivalent.

It is therefore important to extend Hardy's theorem by showing that
universal quantum measurement dynamics requires universal
simultaneity.

It is not feasible to set up the double Bell experiment so that the
curvature of spacetime has a significant and relevant effect, but the
thought experiment is needed to study the effect of the curvature on
the dynamics of quantum measurement.  Just as for flat spacetime, the
condition that there is no backward causality or equivalently that
there is no causal loop \cite{Percival1998b} requires that there is a
time ordering for an event at A with respect to an event
at B, which is spatially separated from the event at A.  Unfortunately
this time ordering depends on the hidden variables or background
field, which are not accessible to current experiments.  Universal
quantum measurement therefore requires a universal time ordering for
events with spacelike separation, which is equivalent to a universal
simultaneity.  This is the Hardy simultaneity theorem, which applies
to curved spacetime.  It is an extension of the original Hardy theorem
and is based on the assumption that the events occur at spacetime
points.  Some latitude is allowed when they take a finite time or
occupy a finite region of space.

Hawking \cite{Hawking1968} and Seifert\cite{Seifert1968} proved that
in all universes in the neighbourhood of which there is no backward
causality, there are timelike functions, spacelike foliations of
spacetime and corresponding definitions of simultaneity
\cite{Tipler1980}.  It is interesting to note that `no backward
causality' is the same condition as that used in \cite{Percival1998b}
to prove that measurement dynamics needs a local Lorentz frame, and is
used here to show that it needs simultaneity.  Hardy showed that
special frames are needed.  However, the example of section 4
illustrates that the existence of a special frame at every point of a
curved spacetime does not imply that there is a consistent definition
of simultaneity.

\vs{\bf 4 Not proper time}

The most obvious choice of a time to define simultaneity is the proper
time of local matter, but this is inconsistent, because, as in
Einstein's experiment, the local matter can have different frames.  In
the neighbourhood of the Earth, we could use the Earth as a standard
of rest, but this also is inconsistent, because in gravitational
fields it leads to `simultaneity' between events that have timelike
separation.

As an example take two small clocks at rest with respect to the
surface of the Earth, one vertically at a height $h$ above the other,
where $h$ is small compared with the radius of the Earth.  They could
be at the top and bottom of a tall building, or one on a table in a
laboratory, and one on the floor beneath it.  Suppose they are
synchronized at time $t=0$ in the rest frame at this time.  At later
times, the general relativistic time dilation due to their
gravitational potential difference is much greater that the special
relativistic time dilation due to their different velocities around
the Earth's centre.  The clocks will show a time difference $\Delta t$
after time $t$, where
\begin{equation} \label{A}
{\Delta t\over t} = {gh\over c^2}
\end{equation}
and $g$ is the acceleration due to the Earth's field near its
surface.

The separation between `simultaneous' events as given by the two
clocks becomes timelike after a time $t$ for which a signal from one
to the other takes a time $\Delta t$, where
\begin{equation} \label{C}
\Delta t = {h\over c},\hx{so that} 
\h10 t={c\over g}\approx 1{\hs1\rm year},
\end{equation}
which is independent of $h$.  The greater height leads to greater time
shifts because of the greater gravitational potential difference, but
the time taken for light to travel between the bodies increases in
proportion.  The fact that $t=c/g$ is so close to the time taken for
the Earth to orbit the Sun is a well-known coincidence.

One could try to define simultaneity by using some kind of average
over local times, but it is not at all clear over what scale the
average should be taken: the scale of the apparatus? of the Earth? of
the Solar System? the galaxy?  or the universe?  An answer to
this question is suggested in section 6.

There is the same problem if we try to use the rest frame of a field
to define simultaneity.  Suppose that we follow Hardy's example (see
also \cite{Stapp1992,Combourieu1993}) of the universal background
radiation.  This provides a local standard of rest.  We could use any
kind of clock in this standard of rest to define the local `time'.
But then there is a gravitational time shift between the clocks in a
gravitational potential well, and outside it.  So again, using these
measures of local time, `simultaneous' events can have a timelike
separation, which is not allowed.  The same applies applies to the
universal background neutrino fields, or any other background field,
of zero or any other rest mass.

Also, as pointed out by Hardy, there is no clear dynamical process
whereby the quantum measurement of individual systems can be made to
depend on the background radiation.

Section 6 shows that the Hawking-Seifert theorem leads to a possible
resolution of this problem of simultaneity.

\vs{\bf 5 Cosmology}

Wheeler \cite{Wheeler1983b} considered the possibility that
entanglement and localization might occur on cosmological
scales.  We have no observational evidence that entanglement survives
over such distances, but the Hardy simultaneity theorem and the 
experiments of Tittel, Gisin et al \cite{Tittel1998a} provide two
strong arguments for the importance of cosmology to quantum
measurement.

According to the Hardy simultaneity theorem, quantum measurement needs
simultaneity between distant events.  There is already a cosmological
definition of simultaneity for the standard models like the
Robertson-Walker metric for isotropic spacetime.  Cosmic standard time
provides a time function, dividing (or foliating) spacetime into
three-dimensional spacelike surfaces of constant cosmic standard time,
which are maximally symmetric subspaces of the whole of the spacetime
\cite{Weinberg1972}.  Any timelike function provides a definition of
simultaneity, in which events with the same functional value are
simultaneous.  However, on scales smaller than the cosmological, the
spacetime of the universe universe in our epoch is not isotropic,
which leads to the problems of defining simultaneity given in the
previous section.

The second follows from the experiments that demonstrate entanglement
over a given distance, which is currently greatest for the Tittel
et. al. experiment.  A system AB consisting of two parts A and B is in
an entangled pure state when AB is in a pure state, but neither A nor
B separately is in a pure state.  Suppose that the systems A and B are
distant from one another.  Then spacelike separated measurements of A
and of B lead to the nonlocality and simultaneity problems of
dynamical theories of measurement.  The experiment of Tittel and his
collaborators \cite{Tittel1998a} shows directly that there can be
entanglement over 10km, so simultaneity must be defined over regions
of this size.  Assuming that generalized measurement is universal, and
that the Earth is typical regarding measurement, following the
Kepler-Galilean principle PR1, entanglement and its destruction by
measurement over distances of 10km is present always and everywhere.
Now fill spacetime with overlapping regions of linear dimension 10km
in space and 10km$/c$ in time.  

For every overlap region the definitions of simultaneity must be
consistent, so by iteration they must be consistent throughout
spacetime, or at least where and when generalized measurement takes
place.  We have no means of checking whether this includes the
neighbourhood of singularities in spacetime or very strong
gravitational fields, including the very early universe, the late
stages of a closed universe, or in the neighbourhood of a black hole,
but the rest of spacetime needs a definition of simultaneity, with a
corresponding timelike function and spacelike foliation.

\vs{\bf 6 Measurement field}


Experimenters in any field of physics who work in a nearly flat
spacetime, and find that the results of their experiments depend on
the Lorentz frame of the apparatus, do not immediately conclude that
special relativity is wrong.  They look for some previously
unsuspected background influence that depends on the environment and
which determines a special frame.  This influence comes from a
background physical system which interacts with the system being
studied.  No one has found a convincing example for which this
procedure fails.  {\it Environment} is used in a broad sense, and may
include fields that penetrate the system.

Similarly, the need for a consistent definition of simultaneity for
measurement does not contradict special relativity.  But it requires a
physical system that defines simultaneity and interacts with the
measured system.  Suppose that this physical system is a {\it
measurement} field $\mu(x)$, where $x=(x^0,x^1,x^2,x^3)$ are the time
and space coordinates of a spacetime point.  Make the following
further assumptions: \vs
\hspace{1cm}\parbox{11cm}{
MU1. For the purposes of the present paper, $\mu(x)$ is a real
classical scalar field.  Later it will have to be quantized, it may
have imaginary components and may not be scalar.

MU2. There was an epoch in the early universe with a a cosmic time
$t$.

MU3. In this early epoch $\mu$ was a monotonically increasing function
of $t$.

MU4. In local inertial frames and in epochs like ours, $\mu(x)$
satisfies the zero-mass Klein-Gordon equation, or wave equation
($c=1$):
\begin{equation} \label{E}
\Big({\de^2\over\de t^2}-\nabla^2\Big)\mu(x) =0.
\end{equation}    
}\vs

From the boundary condition in the early universe, the solution of the
Klein-Gordon equation has a component of zero wavenumber.  In quantum
theory, these solutions are usually ignored.

For an early homogeneous universe, $\mu(x)$ has no space dependence.
It is a timelike function, and the surfaces ${\cal S}_0[\mu(x)=\mu_0]$
form a spacelike foliation of this part of spacetime.  A critical
question for measurement dynamics is whether this is true for all
spacetime.  I have been unable to prove it or find a counterexample,
but there are physical arguments suggesting that in our universe,
sufficiently far from singularities like black holes, the field
$\mu(x)$ is a timelike function.

Locally in inertial frames and globally in flat spacetime, if $\mu$ is
independent of the space coordinates $x^1,x^2,x^3$, then it is a
linear function of $x^0=t$,
$$
\mu = a t + t_0.
$$
$\mu(x)$ is a timelike function in some early epochs, so consider a
spacelike surface ${\cal S}[\mu(x)=\mu_0]$ at such an epoch.  From the
form of the Green function of the Klein-Gordon equation, the solution
at any later spacetime point $x'$ is a weighted mean over the values
and derivatives of $\mu$ on the intersection of the backward light
cone of $x'$ and the surface ${\cal S}$.  If $x'$ is not near a region
of large spacetime singularity, for example if it is on or in the
Earth or the Sun, the local curvature of spacetime is small, and on
scales small compared with the radius of the universe, the light cone
will approximate a light cone for flat space, with the exception of
strong focussing by gravitational lenses, which are rare.  The
weighted mean will then be the mean over a large spherical shell.

According to the cosmological principle, the universe is isotropic,
and therefore homogeneous on sufficiently large scales, so this mean
will approximate the mean for an isotropic universe.  So for such
$x'$, the Green function can be used to propagate the solution forward
in time, as a small perturbation of the solution for an isotropic
universe.  For such a universe, $\mu$ is a timelike function, which
provides the definition of simultaneity needed for quantum measurement.

The required cosmological principle says that on sufficiently large
scales, averages over spherical shells are uniform.  This is a
stronger than the usual cosmological principle, that, usually by
implication, refers to averages over solid spheres or similar regions.

\vs{\bf 7 A simple relativistic measurement dynamics}

A nonrelativistic stochastic measurement dynamics, which is consistent
with the Kepler-Galileo and Newton-Laplace principles, was
proposed long ago by L\"uders \cite{Luders1951}.  

The mathematics comes from the Copenhagen school, particularly von Neumann and
Heisenberg, but the physics is consistent with measurement as a
nonrelativistic dynamical process, not with the Copenhagen
interpretation.  Heisenberg gave a descriptive account in
\cite{Heisenberg1958}.  Gisin used L\"uders dynamics as a starting
point for the quantum state diffusion approach to measurement
\cite{Gisin1984a}.

Because the measurements in this nonrelativistic theory are localized
in spacetime, it can be assumed that they have a definite time order.
Consider just one measurement of a dynamical variable of the quantum
system with corresponding nondegenerate Hermitean operator $\G$ with
eigenstates $|g\>$.  Before the measurement, the quantum system is in
the initial state $|\i\>$ and afterwards it is in a final state
$|\f\>$, which is one of the eigenstates $|g\>$.  The measurement
dynamics is stochastic, and the probability that the system will
finish in state $|g\>$ is
\begin{equation} \label{Q}
\Pr\big(|\f\>=|g\>\big)\hs1 = \<\i|g\>^2.
\end{equation}
In general the classical system also changes its state, from some
initial configuration to a final configuration corresponding to the
measurement of the value $g$ whose probability (\ref{Q}) depends on
$|\i\>$.  This is the influence of the initial quantum state of the
quantum system on the final state of the classical system, that
characterizes generalized quantum measurement.

The stochastic evolution of classical and quantum systems consists of
continuous evolution of each according to their own deterministic
dynamics, with sudden stochastic jumps which correspond to the
measurements that take place at times determined by the classical
system, in which the quantum and classical systems influence each
other.  In this theory, the classical system can influence the quantum
system through time-dependent Hamiltonians whose current value depends
on the state of a classical system.  But the quantum system can only
influence a classical system through a measurement.

In the corresponding relativistic theory, the jumps do not occur at
constant time, but at constant values of $\mu$.  In this way the
measurement field affects the dynamics of quantum measurement, there
are no causal loops, and relativistic principles are preserved,
at least formally.

This picture of measurement is unsatisfactory in several ways.  In
particular, the timing of the jumps is not normally determined by the
classical system alone.  In the modern theory of continuous laboratory
measurements, originating with Davies \cite{Davies1976}, developed by
many authors \cite{Plenio1998,Stenholm1997,Carmichael1993b} and now
used widely in quantum optics, the timing of the jumps is determined
also by the state of the quantum system.  Further, the L\"uders
picture assumes that there are distinctly classical and distinctly
quantal degrees of freedom, with a `shifty split' between them
\cite{Bell1987a}.  This split is convenient for conventional quantum
theory, but there is no evidence that the world is divided in this way
into purely classical and purely quantum domains.

Modern nonrelativistic dynamical theories of measurement have neither
of these problems, but the theory of the measurement field has
not yet been extended to relativistic versions of these theories.

\vs{\bf 8 Discussion and conclusions}

It may seem surprising that tachyons have played no role
\cite{Maudlin1994}.  There is a reason for this.  The usual theory of
tachyons has no preferred frame \cite{Feinberg1967}.  This is normally
considered an advantage, but without the preferred frame, interaction
with normal matter leads to causal loops.  Hence the difficulty of
giving a physical interpretation to such interaction.  Here the
preferred frame is a necessity, through a a nonlocal interaction with
the measurement field, and without using tachyons.

There are many versions of nonrelativistic measurement dynamics
without the faults of the L\"uders theory discussed in section 7.  All
of them are nonlocal, following Bell's theorem.  The first was the
pilot wave theory of de Broglie and Bohm, in which there are both
waves and particles, and measurement dynamics is the result of the
effect of the quantum waves on the classical particles
\cite{Holland1993,Bohm1993}.  Since then there have been dynamical
theories based on waves alone, in which the Schr\"odinger equation 
is modified by a weak stochastic process of localization, leading to
the `collapse' of the quantum wave.  Particles are just very localized
waves \cite{Percival1998c}.  The nonlocal processes depend on
simultaneous changes that take place at spacelike separated points,
where the simultaneity is determined by the universal time variable
$t$.

The theory presented here assumed that events like the orientation of
a polarizer or the recording of a spin take place at spacetime points.
In fact preparation and recording occupy finite regions of space
and take a finite time.  This complicates the theory, but quantum
state diffusion models of nonrelativistic measurement show that this
complication leads to no fundamentally new problems.
\cite{Percival1998c}.

To each of the nonrelativistic dynamical theories there corresponds a
relativistic theory in which the definition of simultaneity is
provided by the measurement field variable $\mu$, just as in the case
of L\"uder's dynamics.  But none of these relativistic theories
represents a complete solution to the problem of relativistic
measurement dynamics.  There are still many unsolved problems.

If there is two-way {\it interaction} between matter and the
measurement field, then superluminal signals might be possible.  Then
conclusions based on their non-existence would no longer hold
\cite{Gisin1989a}.  Their existence leads to no contradiction, because
there is a consistent and universal timelike function which defines
past and future.  In some respects the measurement field plays the
role of an ether, yet there is no conflict with the principle of
special relativity.

Section 6 introduced the measurement field, with the familiar
Klein-Gordon dynamics.  But the dynamics of the interaction between
the measurement field and the matter fields is not familiar: it
introduces the nonlocality of measurement.  Relativistic theories of
measurement dynamics cannot be considered in isolation from the rest
of quantum physics.  The nonlocality of the interaction is not
consistent with the usual local interactions.  For typical quantum
systems studied in laboratory experiments, the nonlocal interactions
have been too weak for their effects to be be seen so far.  But even
now there is a fundamental problem in reconciling the nonlocal
dynamics of measurement with the usual local dynamics of quantum
fields.

It might be that the introduction of nonlocal dynamics could solve
some of the outstanding problems of a `theory of everything' but there
is little evidence for this as yet.  It is unacceptable to have a
dynamical theory of measurement that is incompatible with the theory
of quantum fields or strings, just as it is unacceptable to have a
theory of gravity that has no universally accepted quantization, even
though so far neither the details of measurement dynamics nor quantum
gravity are accessible to experiment.  This paper shows that the
large-scale curvature of spacetime is relevant to the problems of
quantum measurement.  But quantum measurement may or may not be
connected to the problem of quantizing gravity, as has been suggested
\cite{Percival1998c}.

Much remains to be done.

\vs{\bf Acknowledgements} I thank John Charap, Lajos
Di\'osi, Nicolas Gisin, Lucien Hardy, William Power and
Walter Strunz and for helpful communications, and both the
Leverhulme Trust and the UK EPSRC for essential financial support.
I am particularly grateful to George Ellis and Paul Tod for their
help and advice on cosmology.
 
\vskip20mm
\bibliography{../../icp}   

\begin{thebibliography}{10}

\bibitem{Aharanov1981}
Y.~Aharanov and D.~Albert.
\newblock Can we make sense of the measurement process in relativistic quantum
  mechanics?
\newblock {\em Phys. Rev. D}, 24:359--370, 1981.

\bibitem{Aharanov1984a}
Y.~Aharanov and D.~Albert.
\newblock Is the usual notion of time evolution adequate for quantum-mechanical
  systems? I.
\newblock {\em Phys. Rev. D}, 29:223--227, 1984.

\bibitem{Aharanov1984b}
Y.~Aharanov and D.~Albert.
\newblock Is the usual notion of time evolution adequate for quantum-mechanical systems? II.
\newblock {\em Phys. Rev. D}, 29:228--234, 1984.

\bibitem{Aspect1982a}
A.~Aspect, P.~Grangier and G.~Roger.
\newblock Experimental test of Bell inequalities using time-varying analyzers.
\newblock {\em Phys.\ Rev.\ Lett.}, 49:91, 1982.

\bibitem{Aspect1982b}
A.~Aspect, J.~Dalibard and G.~Roger.
\newblock Experimental test of Bell inequalities using time-varying analyzers.
\newblock {\em Phys.\ Rev.\ Lett.}, 49:1804--1807, 1982.

\bibitem{Bell1964}
J.~Bell.
\newblock On the Einstein-Podolsky-Rosen paradox.
\newblock {\em Physics}, 1:195--200, 1964.

\bibitem{Bell1987a}
J.~Bell.
\newblock {\em Speakable and Unspeakable in Quantum Mechanics}.
\newblock Cambridge University Press, Cambridge, England, 1987.

\bibitem{Bell1990a}
J.~Bell.
\newblock Against measurement.
\newblock {\em Physics World}, 3(8):33, 1990.

\bibitem{Bell1990b}
J.~Bell.
\newblock La nouvelle cuisine.
\newblock In A.~Sarlemijn and P.~Kroes, editors, {\em Between Science and
  Technology}, pages 97--115. Elsevier Science (North-Holland), 1990.

\bibitem{Bohm1993}
D.~Bohm and B.~Hiley.
\newblock {\em The Undivided Universe}.
\newblock Routledge, London, 1993.

\bibitem{Bohr1935}
N.~Bohr.
\newblock Can quantum-mechanical description of physical reality be considered
  complete?
\newblock {\em Phys.\ Rev.}, 48:696--702, 1935.

\bibitem{Carmichael1993b}
H.~J. Carmichael.
\newblock {\em An Open Systems Approach to Quantum Optics}.
\newblock Springer, Berlin, 1993.

\bibitem{Clifton1992}
R.~Clifton and P.~Niemann.
\newblock Hardy's theorem for two entangled spin-half particles.
\newblock {\em Phys.\ Lett.\ A}, 166:177--184, 1992.

\bibitem{Combourieu1993}
M.-C. Combourieu and J.-P. Vigier.
\newblock Absolute space-time realism in Lorentz invariant interpretations of
  quantum mechanics.
\newblock {\em Phys.\ Lett.\ A}, 175:269--272, 1993.

\bibitem{Davies1976}
E.~Davies.
\newblock {\em Quantum Theory of Open Systems}.
\newblock Academic, London, 1976.

\bibitem{Einstein1935}
A.~Einstein, B.~Podolsky and N.~Rosen.
\newblock Can quantum-mechanical description of reality be considered complete?
\newblock {\em Phys. Rev.}, 47:777--780, 1935.

\bibitem{Feinberg1967}
G.~Feinberg.
\newblock Possibility of faster-than-light particles.
\newblock {\em Phys. Rev.}, 159:1089--1105, 1967.

\bibitem{Fry1997}
E.~Fry and T.~Walther.
\newblock A Bell inequality experiment based on molecular dissociation.
\newblock In H.~M. Cohen, R.S. and J.~Stachel, editors, {\em Experimental
  Metaphysics}, pages 61--71, Dordrecht, Netherlands, 1997. Kluwer.

\bibitem{Gisin1984a}
N.~Gisin.
\newblock Quantum measurements and stochastic processes.
\newblock {\em Phys.\ Rev.\ Lett.}, 52:1657--1660, 1984.

\bibitem{Gisin1989a}
N.~Gisin.
\newblock Stochastic quantum dynamics and relativity.
\newblock {\em Helvetica Physica Acta}, 62:363--371, 1989.

\bibitem{Hawking1968}
S.~Hawking.
\newblock Existence of cosmic time functions.
\newblock {\em Proc. Roy. Soc. A}, 308:433--435, 1968.

\bibitem{Heisenberg1958}
W.~Heisenberg.
\newblock {\em Physics and Philosophy}.
\newblock Harper and Row, New York, 1958.

\bibitem{Holland1993}
P.~Holland.
\newblock {\em The Quantum Theory of Motion}.
\newblock Cambridge University Press, Cambridge, England, 1993.

\bibitem{Hardy1992a}
L.Hardy.
\newblock Quantum mechanics, local realistic theories and Lorentz-invariant
  realistic theories.
\newblock {\em Phys. Rev. Lett.}, 68:2981--2984, 1992.

\bibitem{Hardy1992c}
L.Hardy.
\newblock Quantum optical experiment to test local realism.
\newblock {\em Phys.\ Lett.\ A}, 167:17--23, 1992.

\bibitem{Luders1951}
G.~L\"uders.
\newblock \"Uber die Zustands\"anderung durch den Messprozess.
\newblock {\em Ann. Phys. (Leipzig)}, 8:322--328, 1951.

\bibitem{Maudlin1994}
T.~Maudlin.
\newblock {\em Quantum Non-locality and Relativity}.
\newblock Blackwell, Oxford, 1994.

\bibitem{Percival1998c}
I.~Percival.
\newblock {\em Quantum State Diffusion}.
\newblock Cambridge University Press, Cambridge, U.K., 1998.

\bibitem{Percival1998b}
I.~Percival.
\newblock Quantum transfer functions, weak nonlocality and relativity.
\newblock {\em Phys.\ Lett.\ A}, 244:495--501, 1998.

\bibitem{Peres1995a}
A.~Peres.
\newblock {\em Quantum Theory: Concepts and Methods}.
\newblock Kluwer Academic Publishers, Dordrecht, The Netherlands, 1995.

\bibitem{Plenio1998}
M.~Plenio and P.~Knight.
\newblock The quantum jump approach to dissipative dynamics in quantum optics.
\newblock {\em Rev.\ Mod.\ Phys.}, 70:101--144, 1998.

\bibitem{Seifert1968}
H.~Seifert.
\newblock {\em Kausal Lorentzr\"aume, Ph.D. Thesis}.
\newblock Hamburg University, Hamburg, 1968.

\bibitem{Shimony1986}
A.~Shimony.
\newblock Events and processes in the quantum world.
\newblock In R.~Penrose and C.~Isham, editors, {\em Quantum Concepts in Space
  and Time}, pages 182--203, Oxford, 1986. Oxford University.

\bibitem{Stapp1992}
H.~Stapp.
\newblock Noise-induced reduction of wave packets and faster-than-light
  influences.
\newblock {\em Phys.\ Rev.\ A}, 46:6860--6868, 1992.

\bibitem{Stenholm1997}
S.~Stenholm and M.~Wilkens.
\newblock Jumps in quantum theory.
\newblock {\em Contemporary Physics}, 38:257--268, 1997.

\bibitem{Tipler1980}
F.~Tipler, C.~Clarke and G.~Ellis.
\newblock Singularities and horizons -- a review article.
\newblock In A.~Held, editor, {\em General Relativity and Gravitation}, pages
  97--206, New York, 1980. Plenum.

\bibitem{Tittel1998a}
W.~Tittel, J.~Brendel, B.~Gisin, T.~Herzog, H.~Zbinden, and N.~Gisin.
\newblock Experimental demonstration of quantum correlations over more than 10
  km.
\newblock {\em Phys.\ Rev.\ A}, 57:3229--3232, 1998.

\bibitem{Tittel1998b}
W.~Tittel, J.~Brendel, H.~Zbinden, and N.~Gisin.
\newblock Violation of Bell inequalities by photons more than 10km apart.
\newblock {\em Phys. Rev. Lett.}, 81:3563--3566, 1998.

\bibitem{Walther1997}
T.~Walther and E.~Fry.
\newblock On some aspects of an Hg based EPR experiment.
\newblock {\em Z. f. Naturforschung Section A-A J. of Phys. Sci.}, 52:20--24,
  1997.

\bibitem{Weinberg1972}
S.~Weinberg.
\newblock {\em Gravitation and Cosmology}.
\newblock Wiley, New York, 1972.

\bibitem{Wheeler1983b}
J.~Wheeler.
\newblock Law without law.
\newblock In J.~Wheeler and H.~Zurek, editors, {\em Quantum Theory and
  Measurement}, pages 182--213, Princeton, N.J., 1983. Princeton University.

\bibitem{Wheeler1983a}
J.~Wheeler and H.~Zurek, editors.
\newblock {\em Quantum Theory and Measurement}.
\newblock Princeton University, Princeton, N.J., 1983.

\end{thebibliography}
\bibliographystyle{abbrv}   

\end{document}